\documentclass[%
 reprint,
 amsmath,amssymb,
 aps,
longbibliography,
floatfix
]{revtex4-2}

\usepackage{graphicx}
\usepackage{dcolumn}
\usepackage{bm}
\usepackage{xcolor}
\usepackage{overpic}
\usepackage{hyperref}

\usepackage{rotating}
\usepackage{tikz}
\usetikzlibrary{arrows.meta}

\newcommand{\unitz}{\hat{\boldsymbol e}_z}

\newcommand{\unitn}{\hat{\boldsymbol e}_n}
\newcommand{\units}{\hat{\boldsymbol e}_s}

\newcommand{\vel}{\boldsymbol u}

\newcommand{\up}{\boldsymbol u_\perp}

\newcommand{\bup}{\widetilde{\boldsymbol u}_\perp}

\newcommand{\cur}{\boldsymbol c}

\renewcommand{\cp}{\boldsymbol c_\perp}

\newcommand{\bcur}{\widetilde{\boldsymbol c}}

\newcommand{\bcp}{\widetilde{\boldsymbol c}_\perp}

\newcommand{\cwb}{c_w^-}
\newcommand{\cwt}{c_w^+}
\newcommand{\cwtb}{c_w^\pm}

\newcommand{\J}{\boldsymbol J}

\newcommand{\Jp}{\boldsymbol J_\perp}

\newcommand{\bJp}{\widetilde{\boldsymbol J}_\perp}

\newcommand{\nabp}{\nabla_\perp}

\newcommand{\order}{\mathcal O}

\newcommand{\Ro}{R\!o}
\newcommand{\Ek}{E\!k}

\newcommand{\Rm}{R\!m}

\newcommand{\zbot}{z^{\rm bot}}
\newcommand{\ztop}{z^{\rm top}}

\newcommand{\ordert}[1]{\mathcal O\!(#1)}

\begin{document}

\preprint{APS/123-QED}

\title{Magnetic Taylor-Proudman Constraint Explains Flows into the Tangent Cylinder}
\author{Alban Poth\'erat}
\email{alban.potherat@coventry.ac.uk}
\affiliation{Coventry University, Centre for Fluid and Complex Systems, Mile Lane, Coventry   CV1 2NL, UK}
\author{K\'elig Aujogue}
\affiliation{Coventry University, Centre for Fluid and Complex Systems, Mile Lane, Coventry CV1 2NL, UK}
\author{Fran\c cois Debray}
\affiliation{Laboratoire National des Champs Magn\'etiques Intenses (LNCMI), CNRS UPR 3228, EMFL, Universit\'e Toulouse III - Paul Sabatier, Universit\'e F\'ed\'erale Toulouse Midi-Pyr\'en\'ees, Institut National des Sciences Appliqu\'ees, Universit\'e Grenoble Alpes, 38042 Grenoble CEDEX, France}



%

\date{\today}

\begin{abstract}

Tangent Cylinders (TCs) have shaped our understanding of planetary dynamos and liquid cores. 
The Taylor-Proudman Constraint (TPC) due to planetary rotation creates these imaginary surfaces separating polar and equatorial regions but cannot explain the flows meandering through them. %
Here we establish and verify experimentally that magnetic fields aligned with rotation %
drive flows \emph{into} TCs, linked to the flows \emph{along} TCs by a \emph{magnetic} Taylor-Proudman constraint. %
This constraint %
explains and quantifies how magnetic fields reshape rotating flows in planetary interiors and magnetorotating flows in general. %
\end{abstract}

\maketitle
\begin{figure}[h!]%
	\centering%
	\includegraphics[width=0.8\columnwidth]{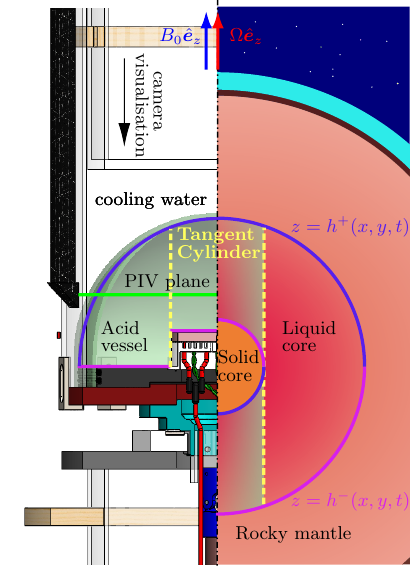}
	\caption{Sketch of the Earth's structure with liquid outer core and solid inner core (right) and the Little Earth Experiment, its rotating hemisphere filled with sulphuric acid, protruding heater, cooling tank and height-adjustable PIV system (left). 
\label{fig:lee}}%
\end{figure}

Geophysical and astrophysical flows, whether atmospheres, accretion disks, stellar or planetary interiors are often strongly constrained by the Coriolis force produced by background rotation. The Taylor-Proudman Constraint (TPC) expresses that rotation mainly renders the flow 2D, and 2D-solenoidal in the plane normal to rotation 
\cite{proudman1916_prsa,taylor1917_prsa,greenspan1969}. To remain 2D and 2D-solenoidal in planetary interiors shaped as spherical shells, the flow must follow geostrophic contours, aligned with surfaces of constant height, \emph{i.e.} concentric cylinders \cite{greenspan1969}.
For this reason, in planetary interiors with a solid inner core, the "tangent cylinder" (TC, Fig. \ref{fig:lee}) extruded from the solid core along the rotation axis, is often considered as a mechanical barrier separating polar and equatorial regions. Both regions are convective, but with different dynamics and they are believed to play different roles in the dynamo mechanism that sustains planetary magnetic fields 
\cite{takahashi2003_pepi,sreenivasan2006_gafd,garcia2008_prl,schaeffer2017_gji,gastine2023_jfm}.\\
The paradox, here is that many dynamo theories, especially for the Earth, demand that the Lorentz and the Coriolis forces be of the same order, \emph{i.e.} in the \emph{magnetostrophic regime}, to recover large-scale magnetic fields \cite{dormy2016_jfm,horn2022_prsa}.
Within the TC too, the TPC determines the dynamics of convective Taylor columns \cite{grooms2010_prl,kunnen2021_jot,ecke2023_arfm}, and the Lorentz force needs to act at the leading order to recover the large structures conducive to dynamo action there \cite{chandrasekhar1961, eltayeb1972_prsa, eltayeb1975_jfm,roberts2013_prp,aps2015_pf,horn2022_prsa,horn2023_prsa}. 
However, where the Lorentz forces competes with the Coriolis force, the TPC is violated 
\cite{cao2018_pnas,hotta2018_apjl,sakuraba2002_gafd,mason2022_gafd}. 
The scenario of a flow structure articulated around an impermeable TC then breaks down as it cannot explain recently discovered flows meandering in and out of it \cite{schaeffer2017_gji,finlay2023_nat}.
Interestingly, magnetohydrodynamic (MHD) flows in a homogeneous magnetic field exhibit similar tendencies to two-dimensionality to rotating flows\cite{sm82, kolesnikov1974_ian, kp2010_prl, bpdd2018_prl}, and also follow geostrophiclike contours aligned with so-called "characteristic surfaces" 
\cite{hunt1968_jfm3,kulikovskii1973_ian,alboussiere1996_pf}. Yet, although both axial magnetic fields and rotation act individually {alike}, 
a theory for their combined action is both missing and needed to understand the flow in the TC region, and how Lorentz and Coriolis forces constrain the convective structures inside it. Here we derive and experimentally verify the physical foundation of such a theory.\\ 

The physical model is a fluid (density $\rho$, viscosity $\nu$, electric conductivity $\sigma$) with boundaries at $z=h^-(x,y,t)$ and $z=h^+(x,y,t)$, subject to a magnetic field $\boldsymbol B(x,y,z,t)$, assumed 
along $\unitz$ for simplicity, and to a constant background rotation $\Omega\unitz$. 
Splitting vectors and operators into their components along and normal to $\unitz$ (subscripts $z$ and $_\perp$), \emph{e.g.} $\nabp=(\partial_x,\partial_y)$, the dimensionless velocity $\vel$, pressure $p$ and electric current $\J$ are governed by the Navier-Stokes equations
\begin{eqnarray}%
\!\Ro(\!\partial_t\!+\!\vel\!\cdot\!\nabla\!)\up\!+\!\nabp p\!&\!=\!&\!(\!\up\!+\!\Lambda\Jp B_z\!)\!\times\!\unitz\!+\!\Ek\Delta \up\!, 
	\label{eq:nsp}\\%
        \Ro(\!\partial_t\!+\!\vel\!\cdot\!\nabla\!)u_z\!+\!\partial_z p\!&\!=\!&\!f_b+\Ek\Delta u_z,%
        \label{eq:nsz}\\%
        \nabla\cdot\vel&=&0.%
        \label{eq:contu}%
\end{eqnarray}%
The Ekman, Rossby numbers $\Ek=\nu/(2\Omega h_0^2)$, $\Ro=U/(2\Omega h_0)$,
and buoyancy force $f_b=\ordert{\Ro}$\cite{aurnou2020_prf} 
based on typical velocity $U$ 
and outer core shell thickness $h_0$ are very small in the regime of fast rotation relevant to planets (for the Earth $\Ek\simeq10^{-15}$ and $\Ro\gtrsim 10^{-6}$), but the Elsasser number $\Lambda=\sigma B_0^2/(2\rho \Omega)\in [10^{-1},10^2]$ based on typical axial field $B_0$ is not
\cite{schubert2011_pepi}. 
The electric current $\J$ is expressed by Ohm's law in terms of the electric potential $\phi$ and magnetic vector-potential $\boldsymbol A$, and satisfies charge conservation,
\begin{eqnarray} %
	\J&=&-\nabla \phi+{B_z}\vel\times\unitz+\partial_t\boldsymbol A,%
        \label{eq:qsmhd_ohm} 
        \qquad \nabla\cdot\J=0%
        \label{eq:qsmhd_contj}.%
\end{eqnarray}%
In geophysical context, these equations are usually 
closed with the induction equation and magnetic flux conservation.
The magnetic TPC, however, only requires an expression of the electric current.
The idea 
is to express the total Lorentz-Coriolis force 
as the product of a single {solenoidal current $\cur$
and $\unitz$. 
$\cur$ is chosen as the compound current of mass and charge 
$\cp=\up+\Lambda B_z\Jp$ and its vertical component $c_z=u_z+\Lambda (B_zJ_z-\int\J\cdot\nabla B_zdz)$ is defined up to an integration constant to ensure conservation of the compound massic and electric charge associated to $\cur$, \emph{i.e.}  $\nabla\cdot\cur=0$.}
Then at $\order{(\Ek,\Ro)}$, Eqs. (\ref{eq:nsp}), ({\ref{eq:nsz}}) imply that bulk variables  (denoted $\widetilde{\cdot}$) satisfy the "magnetic Taylor-Proudman constraint" (MTPC) \cite{sakuraba2002_gafd,mason2022_gafd}, 
\begin{eqnarray}%
	\partial_z\bcp&=&\partial_z{\bup}+\Lambda\partial_z(B_z\bJp)=0,%
\label{eq:cp_bulk}\\%
	\nabp\cdot\bcp&=&\nabp\cdot\bup+\Lambda\nabp\cdot (B_z\bJp)=0.%
\label{eq:mtp}%
\end{eqnarray}%
These equations express that in the rapidly rotating limit, 
the \emph{current} becomes quasi-2D and quasihorizontally solenoidal. 
Without magnetic field, the bulk horizontal velocity $\bup$ becomes 2D and horizontally solenoidal. Here, by contrast, $\bup$ may be $z$-dependent and horizontally divergent, as long as $\Lambda\bJp B_z$ possesses the opposite $z$-dependence and horizontal divergence.\\ 
The magnetic counterpart of the topological constraint forcing the flow along geostrophic contours 
is found by integrating the conservation of compound charge ($\nabla\cdot\cur=0$) along $\unitz$. {Noting that regardless of the 
boundary conditions at $z=h^\pm$,
the current \emph{at} each boundary $c_z(z=h^\pm)$ is the sum of the current due to the motion of the boundaries (neglecting the electric displacement current) and the current \emph{through} them 
$\cwtb$},  
\begin{equation}%
	c_z(z=h^\pm)=\partial_th^\pm+\cp^\pm\cdot\nabp h^\pm+\cwtb, %
	\label{eq:surface_condition}%
\end{equation}%
it follows that
\begin{equation}%
	\int_{h^-}^{h^+} \nabp\cdot\cp dz =c_w{-\partial_th-\cp^+\cdot\nabp h^++\cp^-\cdot\nabp h^-},%
\label{eq:cont_int}%
\end{equation}%
where $h=h^+-h^-$ and $c_w=\cwb-\cwt$.
To account for the viscous boundary layers of thickness $\delta\sim\Ek^{1/2}$ \cite{acheson1973_arfm} at $z=h^\pm$,
we decompose the current $\cur$ into its bulk value $\tilde\cur$ satisfying Eqs. (\ref{eq:cp_bulk}), (\ref{eq:mtp}) and 
two corrections for the boundary layers $\boldsymbol c^{\rm BL^\pm}-\widetilde{\boldsymbol c}$, which vanish in the bulk,
\begin{equation}%
	\lim_{|z-h^\pm|\delta^{-1}\rightarrow\infty} \cur^{{\rm B  L}^\pm}-\bcur=0.%
\end{equation}%
Then, using the Leibniz integral rule, 
\begin{eqnarray}%
\int_{h^-}^{h^+} \nabp\cdot{\cp} dz&=& %
\bcp\cdot\nabp\int_{h^-}^{h^  +} dz%
	\label{eq:divc_expanded}\\%
	+ (\nabp\cdot\bcp)\int_{h^-}^{h^+}dz %
&-&\cp^+\cdot\nabp h^+%
+\cp^-\cdot\nabp {h^-}%
\nonumber \\%
+\nabp\cdot\int_{h^-}^{h^+} {\cur^{{\rm BL}^-}_\perp}-\bcp dz%
&+&\nabp\cdot\int_{h^-}^{h^+} {\cur^{{\rm BL}^+}_\perp}-\bcp dz.%
\nonumber%
\end{eqnarray}%
The second term in the \emph{rhs} 
is $\mathcal O(\Ek,\Ro)$ due to the MTPC (\ref{eq:mtp}). The last two terms stem from the current generated within the boundary layers to balance viscous forces, and fed into the bulk. From the compound charge conservation in the boundary layer, these scale 
 as $c_\perp \delta=\ordert{Ek^{1/2}}$. Then from (\ref{eq:cont_int}) and (\ref{eq:divc_expanded}), the bulk current obeys the "topological MTPC":
\begin{equation}%
      \bcp\cdot \nabp h =c_0+\order{(\Ek^{1/2},\Ro)},%
          \label{eq:geostrophic_contours}%
\end{equation}%
where $c_0=-\partial_th+c_w$.  
Equation (\ref{eq:geostrophic_contours}) generalizes the concept of geostrophic contour to include not only 
the effect of the Lorentz force but also permeable, electrically conducting and moving boundaries:
when boundaries are impermeable to $\cur$ ($c_w=0$) and steady ($\partial_th=0$),
it constrains the quasi-2D current $\bcp$ to follow the geostrophic contours. The current may however deflect from them to absorb current ($c_0$) either incurred by the motion of the boundaries (${-\partial_th}$) or fed through them ($c_w$)
if {they} are not impermeable or not electrically insulating.\\ 
We can now express the MTPC in terms of the velocity field: Eq. (\ref{eq:geostrophic_contours}) 
expresses the geometric constraint on a quasi-2D solenoidal current in a 3D domain, regardless of the definition of that current, 
so that Eq. (\ref{eq:geostrophic_contours}) and Ohm's law (\ref{eq:qsmhd_ohm}) impose independent constraints 
from which $\J$ can be eliminated. 
Denoting normal and tangential components to a given geostrophic contour {in the $z$-averaged plane} by indices $n$ and $s$,  and $u_0=c_0(\partial_nh)^{-1}$, these imply (dropping tildes)
\begin{eqnarray}%
	u_n+\Lambda B_z J_n&=&u_0+\ordert{\Ek^{1/2},\Ro},%
\label{eq:un} \\%
	J_n&=&-\nabla_n\phi+B_zu_s+\partial_t A_n.%
\label{eq:jn}%
\end{eqnarray}%
Using electric charge conservation 
and Ohm's law (\ref{eq:qsmhd_ohm}), and 
considering time-averaged quantities (denoted $\langle\cdot\rangle_t$),  (\ref{eq:un}) and (\ref{eq:jn}) yield the "kinematic MTPC",
\begin{eqnarray}%
	\Delta \langle B_z^{-1}(u_n-u_0)\rangle_t&& \nonumber\\%
	+\Lambda[{\nabla}({\nabla}\cdot\langle B_zu_n\rangle_t\units)\cdot\unitn&+&\Delta_{zs}\langle B_zu_s\rangle_t]=0.%
\label{eq:kinematic_mtp_ns}%
\end{eqnarray}%
The key result here is that unlike the current $\cp$, the flow $\up$ does not follow the
 geostrophic contours, even if $c_0=0$. Instead, it is constrained by Eq. (\ref{eq:kinematic_mtp_ns}) to cross them. In planetary interiors with a solid core, flows across TCs have been observed or simulated but never quantified
 \cite{cao2018_pnas,mason2022_gafd,finlay2023_nat}. The kinetic MTPC (\ref{eq:kinematic_mtp_ns}) can measure which part of these through-flows is incurred by the axial component of the planetary magnetic field\ref{fnote:b}.\\

We shall now seek experimental evidence for these flows and for their compliance to the MTPC
in a simple TC geometry. 
%
%
Experiments are performed in the "Little Earth Experiment" (LEE) device \cite{apbds2016_rsi,apsd2018_jfm,aujogue2016_phd}. 
LEE is a model for the convection in the TC 
subject to background rotation and externally imposed axial magnetic field $B=B_0\unitz$. 
The working fluid is sulphuric acid at 30\% mass (viscosity $\nu=1.3\times10^{6}$ m$^2$/s, density $\rho=1.3$ kg/m$^3$, thermal expansion coefficient $\alpha=5.5\times10^{-4}$ K$^{-1}${, thermal diffusivity $\kappa=1.7\times10^{-7}$ m$^2$/s and Prandtl number $\Pr=12$)}. 
Its transparency and electric conductivity ($\sigma=83\pm2 $ Sm)
make it the ideal choice for optical velocimetry in MHD flows \cite{andreev2013_jfm,mph2020_ef}. 
MHD experiments are usually conducted with liquid metals, with an electrical conductivity {$\sim10^4$} higher than sulfuric acid. Hence, to keep $\Lambda=\sigma B_0^2/(2\rho\Omega)\sim 1$ with $\Ek\sim10^{-5}$, 
 magnetic fields $\sim 10$ T are needed, \emph{i.e.} $\sim10^2$ times greater than in liquid metal convection experiments \cite{grannan2022_jfm,vogt2021_jfm}.\\ 
At LEE's heart lies a hemispherical transparent vessel filled with the acid (inner diameter $2R=0.285$ m), 
{rotating} around its vertical axis (Fig. \ref{fig:lee}). It is placed at the bottom of a cylindrical tank filled with water that ensures a constant, cold temperature $T_C$ at all its boundaries but the bottom one. At the bottom, a coaxial cylindrical heating element (radius {$R_{\rm TC}=0.05$} m) protruding by $h_H=0.0225$ m into the vessel fulfils two functions: first, its upper surface is kept at fixed hot temperature $T_H$ to create an unstable temperature gradient prone to drive convection within the vessel. Second, its edge incurs a radial jump in domain height akin to the equatorial edge of the Earth's inner core, where a TC develops under fast rotation.
The heater thus acts as a solid core and the hemispherical vessel wall represents the core-mantle boundary. The heater side wall and the vessel bottom wall are thermally insulated.\\
\noindent In acid, the magnetic diffusivity $\eta$ 
is much greater than in metals,
so for typical velocities of $10^{-3}$ to $10^{-1}$ ms$^{-1}$, 
the magnetic Reynolds number $\Rm=Uh_0/\eta{=U(R-h_H)/\eta}$ measuring the ratio of magnetic field advection to magnetic diffusion is, at most around $10^{-6}$. This places LEE in the quasistatic MHD (QSMHD) regime for which the induced magnetic field 
 is negligible compared to the externally imposed magnetic field $B_0$, so 
$B_z=1$ \cite{roberts1967}.\\
Particle Image Velocimetry (PIV) provides two time-dependent velocity components in 
two planes: at {$z^{\rm bot}=0.22h_0=0.24h_{\rm TC}$}, above the heater and {$z^{\rm top}=0.74h_0=0.8h_{\rm TC}$}, just below the point where the TC meets the inner vessel wall {($h_{\rm TC}$ is the TC's height at $r=R_{\rm TC}$)}.
The temperature is measured at the heater surface and at the outer surface of the acid vessel wall with thermocouples, and {time-averaged} to calculate the free-fall velocity $U=[g\alpha(T_H-T_C)h_0]^{1/2}$ and $\Ro$\cite{aurnou2020_prf}.
Velocity fields are then recorded over 10 min when the difference between these temperatures is statistically steady. All data presented here is averaged over that time.\\ 
\begin{figure}[hpt!]%
\centering%
	\includegraphics[width=\columnwidth]{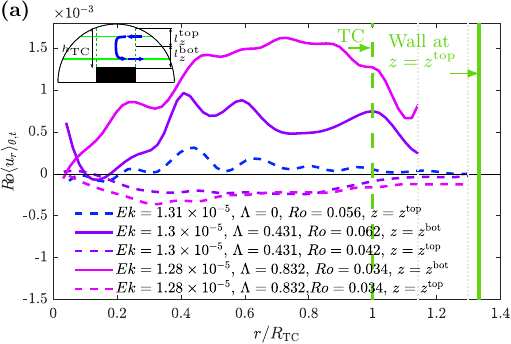}	
	\includegraphics[width=\columnwidth]{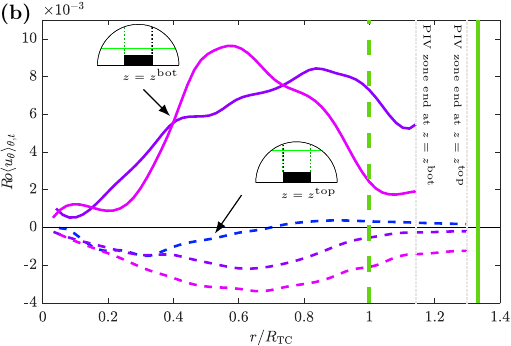}	
	\caption{Time and azimuthally averaged radial profiles of radial (a) and azimuthal (b) velocities in LEE, renormalized by rotation speed, which is more precisely measured in LEE than $U$. 
	The inset sketches show (a) the single cell recirculation and (a), (b) the positions of the PIV planes.
\label{fig:velocity_profiles}}%
\end{figure}

The TC in LEE provides a geostrophic contour to test the relation predicted by the MTPC between the time-averaged azimuthal flow 
along it and radial flow through it without boundary current ($c_0=0$) [indices $(s,n)$ become $(\theta,r)$ in Eq. (\ref{eq:kinematic_mtp_ns})]. 
While in the absence of magnetic field, the classical TPC extinguishes the flow through the TC when  $\Ek\ll1$ and $\Ro\ll1$ \cite{aurnou2003_epsl,apsd2018_jfm}, the MTPC predicts that any azimuthal flow with a nonlinear $z$-dependence should incur a radial flow for $\Lambda>0$.
This flow is tracked through the 
azimuthally and time-averaged profiles of both velocity components in both PIV planes.   
The examples plotted in Fig. \ref{fig:velocity_profiles} 
for $\Ro\in[0.034,0.062]$, $\Ek\simeq 1.3\times10^{-5}$ and $\Lambda$ from 0 to 0.832, span the geostrophic and magnetostrophic regimes where the MTPC applies. 
Magnetic ($\Lambda>0$) and nonmagnetic ($\Lambda=0$) cases exhibit very different velocity profiles: 
while a significant azimuthal flow exists in all cases (in the nonmagnetic case, it is due to the thermal wind incurred by the radial temperature gradient \cite{apsd2018_jfm}); the nonmagnetic flow only has a nonzero radial component in localized convective plumes within the TC ($r/R_{\rm TC}<1$), visible through radial oscillations in the profile. The radial velocity also remains 0 to within measurement precision at the TC boundary. The thermal plumes and corresponding oscillations subsist for $\Lambda>0$ in the top plane. In stark contrast to the nonmagnetic case, however, a radially converging inflow occupies the entire TC in the top plane.
In the bottom plane, a radial flow also appears for $\Lambda>0$ but outward. Hence, the patterns of radial velocity on both planes imply that a meridional recirculation made of at least one cell crosses the TC boundary when $\Lambda>0$. Furthermore, the intensity of this recirculation relative to the azimuthal flow increases with $\Lambda$,  
and is accompanied by a much stronger zonal flow than for $\Lambda=0$. This flow structure subsists at all background rotations ($\Ek\in\{0.82,1.3,3\}\times10^{-5}$) and forcing parameters ($5\times10^{-4}\leq \Ro\leq 0.3$) we explored, 
regardless of variations in flow topology.
Both the meridional flow pattern and the enhanced zonal flow are consistent with the relation between radial and azimuthal flow predicted by the MTPC (\ref{eq:kinematic_mtp_ns}). We shall now quantify this relation across LEE's parameter range.\\
%
%
\begin{figure}[h!]%
	\includegraphics[width=\columnwidth]{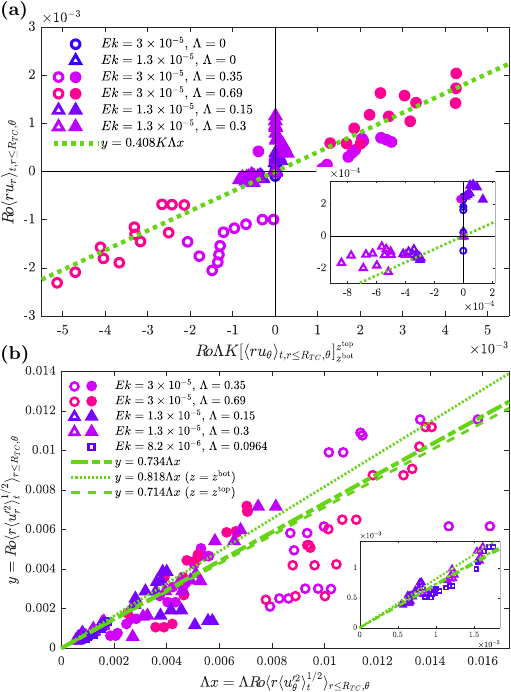}
	\caption{Radial \emph{vs.} azimuthal velocities, azimuthally and radially averaged within the TC (a) time-averaged velocities (b) \emph{rms} of velocity fluctuations, and fit to corresponding scaling relations derived from the MTPC. {Full (open) symbols: measurements at $z=\zbot$ ($z=\ztop$). Different markers of the same type correspond to different thermal forcing, spanning $5\times10^{-4}\leq\Ro\leq0.3$ across all cases.}
\label{fig:mtp_exp}}%
\end{figure}%

LEE's 2D PIV data cannot give the derivatives of the velocity in both axial and radial directions that a local evaluation of the MTPC terms would demand. Hence the MTPC is averaged azimuthally, {and radially within the TC} to yield a scaling relation more amenable to experimental comparison {(see Supplementary Material \footnote{See Supplemental Material for the method to calculate $K(z)$ in Eq. (\ref{eq:mtpc_scaling}) from experimental data}):
\begin{equation}%
	\langle r u_r(z)\rangle_{t,\theta,r\leq R_{\rm TC}}\sim K(z)[\Lambda \langle ru_\theta\rangle_{t,\theta,r\leq R_{\rm TC}}]^{z^{\rm top}}_{z^{\rm bot}}.%
\label{eq:mtpc_scaling}%
\end{equation}%
Coefficient $K(z)$ is obtained in each PIV measurement plane at $z=\zbot$ and $z=\ztop$ from measurements in both planes, by estimating derivatives in Eq. (\ref{eq:kinematic_mtp_ns}) as the ratio of velocities averaged {over horizontal sections of the TC} to the length scales of velocity gradients:
the length scale for $\partial_{zz}^2u_\theta$ between the two PIV planes is the distance between them.}
For $\partial_{zz}^2u_r$, the length scales near the top and bottom planes may differ and are constrained by mass conservation, assuming a single meridional recirculation cell.\\
Both sides of Eq. (\ref{eq:mtpc_scaling}) are shown in Fig. \ref{fig:mtp_exp}(a).
The data collapses approximately along {$\langle r u_r\rangle_{t,r\leq R_{\rm TC},\theta}= 0.408 K\Lambda  [\langle r u_\theta\rangle_{t,r\leq R_{\rm TC},\theta}]_{\zbot}^{\ztop}$}(fitted with confidence interval 0.95), as predicted by the MTPC scaling for the mean flow (\ref{eq:mtpc_scaling}). 
The concentration around the origin ($u_r\simeq0$) of all points at $\Lambda=0$ confirms that the flow across the TC is controlled by the classical TPC in the absence of magnetic field, so that
averages within the TC provide a good measure of the flow across it. Radial averaging still incurs some data scattering, as flows recirculating within the TC (\emph{e.g} through convective plumes) add a contribution to $\langle r u_r\rangle_{t,r\leq R_{TC},\theta}$ that is not constrained by the MTPC. A measure of this effect is captured by the nonzero values of $\langle r u_r\rangle_{t,r\leq R_{TC},\theta}$ for $\Lambda=0$ visible in the inset: {it introduces a greater relative error where either the velocities or $\Lambda$ are small. Nevertheless,} the linear trend imposed by the MTPC clearly dominates.\\
Lastly, the fact that LEE operates in the QSMHD regime offers additional insight into the MTPC: since
$\partial_t A\sim \Rm |\vel\times\unitz|$, it can be neglected in Eq. (\ref{eq:qsmhd_ohm}) so that Eq. (\ref{eq:kinematic_mtp_ns}) applies at each time $t$, instead of only on average. Hence, in QSMHD regimes, the MTPC applies to velocity fluctuations $\vel^\prime$ too. Azimuthal averages of \emph{rms} of velocity fluctuations are, unlike the time-averaged velocities, not constrained by global mass conservation, so 
the gradients of $u_r^\prime$ and $u_\theta^\prime$ share similar length scales: data from both planes 
collapse on almost the same fitted line {$\langle ru_r^{\prime 2}\rangle_{t,r\leq R_{\rm TC},\theta}^{1/2}=0.734\Lambda \langle ru_\theta^{\prime 2}\rangle_{t,r\leq R_{\rm TC},\theta}^{1/2}$}, shown in Fig. \ref{fig:mtp_exp} (b), with similar slopes for top and bottom planes.
This shows not only that the average flow through the TC is controlled by the MTPC, but that so are its fluctuations within the QSMHD regime.\\ 
%
These results show that when magnetic field and background rotation are aligned, the MTPC reshapes the global structure of flows near geostrophic contours such as planetary TCs. The MTPC provides a quantitative tool to predict the topology of such flows dominated by competing Coriolis and Lorentz forces. 
Indeed, the evidence in LEE that a polar magnetic field incurs a meridional flow in the polar region, but also significantly enhances the zonal flow there provides quantitative theoretical support to recent dynamo simulations and observations where these phenomena occur \cite{schaeffer2017_gji,cao2018_pnas,finlay2023_nat}. 
The MTPC also offers a theoretical framework to extend the current
understanding of how the TPC shapes convective Taylor columns in rotating convection \cite{grooms2010_prl,nieves2014_pf} to
similar structures observed in magnetorotating convection \cite{yan2019_jfm,horn2022_prsa,horn2023_prsa}.
In actual planetary interiors, however, the MTPC only applies in regions of axial magnetic field (Eqs. (\ref{eq:cp_bulk}), (\ref{eq:mtp}))
and so may {need amending 
to capture the topology of}
meridional TC flows incorporating the magnetic field topology
\footnote{Nonaxial fields could be incorporated in the MTPC following the steps leading to Eq. (\ref{eq:kinematic_mtp_ns}), as long as the axial Lorentz force is at most $\ordert{\Ro}$ \label{fnote:b}}. 
Similarly, with the generalized MTPC  (11) incorporating conducting, permeable and moving boundaries, more complex dynamics 
at the solid-liquid core boundary, such as unsteadiness due to the solidification-remelting cycle of the inner core \cite{alboussiere2009_nat}, or stratification \cite{juarez2009_qjrms} may also be accounted for.
Finally, the MTPC applies to magnetorotating flows in general, so 
 its implications beyond geophysics and astrophysics also concern numerous other fields such as material science, \emph{e.g.} alloy casting and stirring or crystal growth \cite{grants2017_jcg}.
\begin{acknowledgments}
	The authors are grateful to the European Magnetic Field Laboratory (EMFL) and the Laboratoire des Champs Magn\'etiques Intense-Grenoble (CNRS) for support and access to unique magnets with high magnetic field in sufficiently large bores to conduct fluid mechanics experiments. Subscription to EMFL is funded by EPSRC grant no. NS/A000060/1. This work was supported by Leverhulme Trust grants no. RPG-2012-456 and no. RPG-2017-366, {and EPSRC grant no. EP/X010937/1}. No Artificial Intelligence technology was used in this work, besides spell-checking.%
\end{acknowledgments}
\bibliography{pad2024_prl}

\newpage
\pagestyle{empty}
\begin{figure*}
\includegraphics{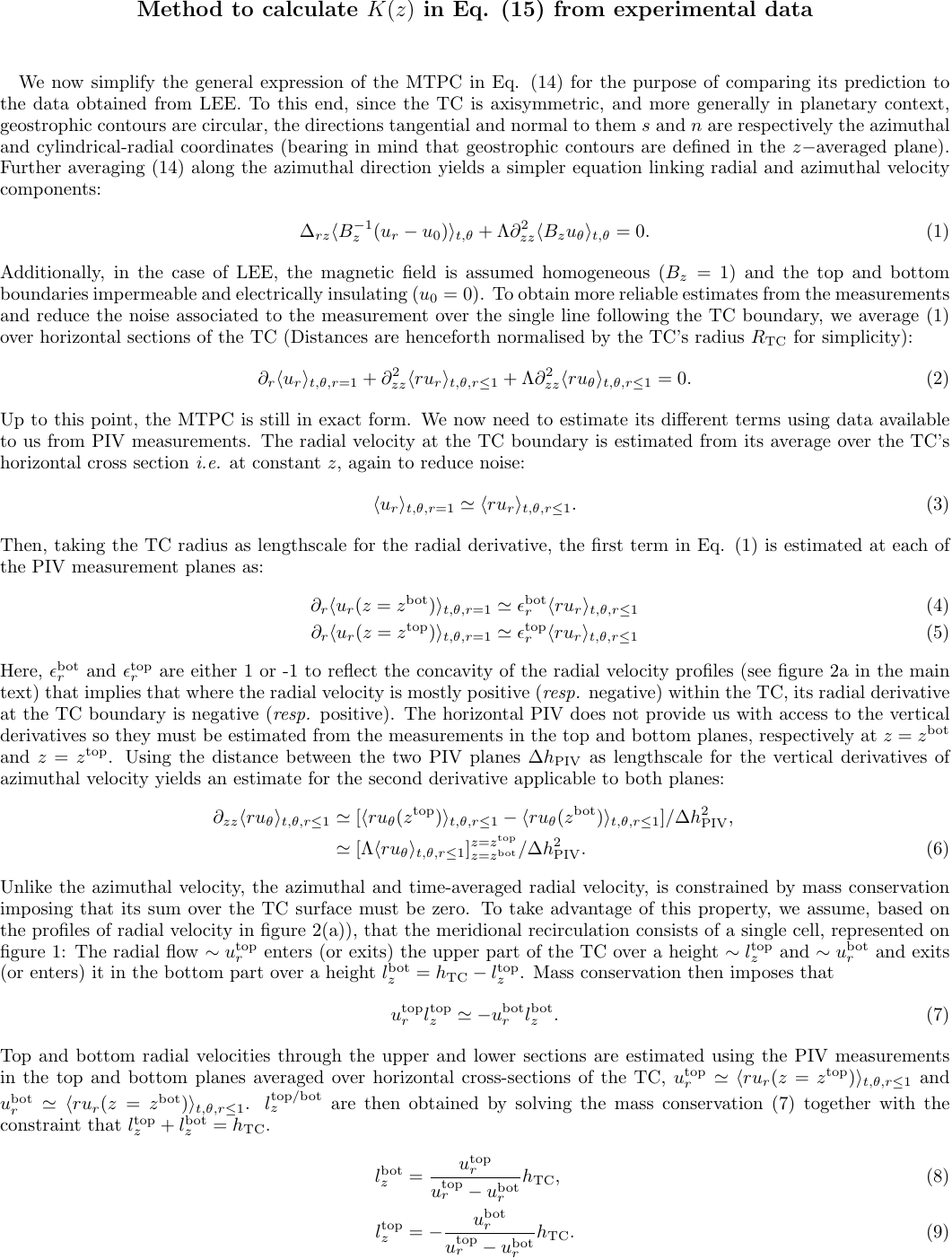}
\newpage
\end{figure*}
\begin{figure*}
\includegraphics{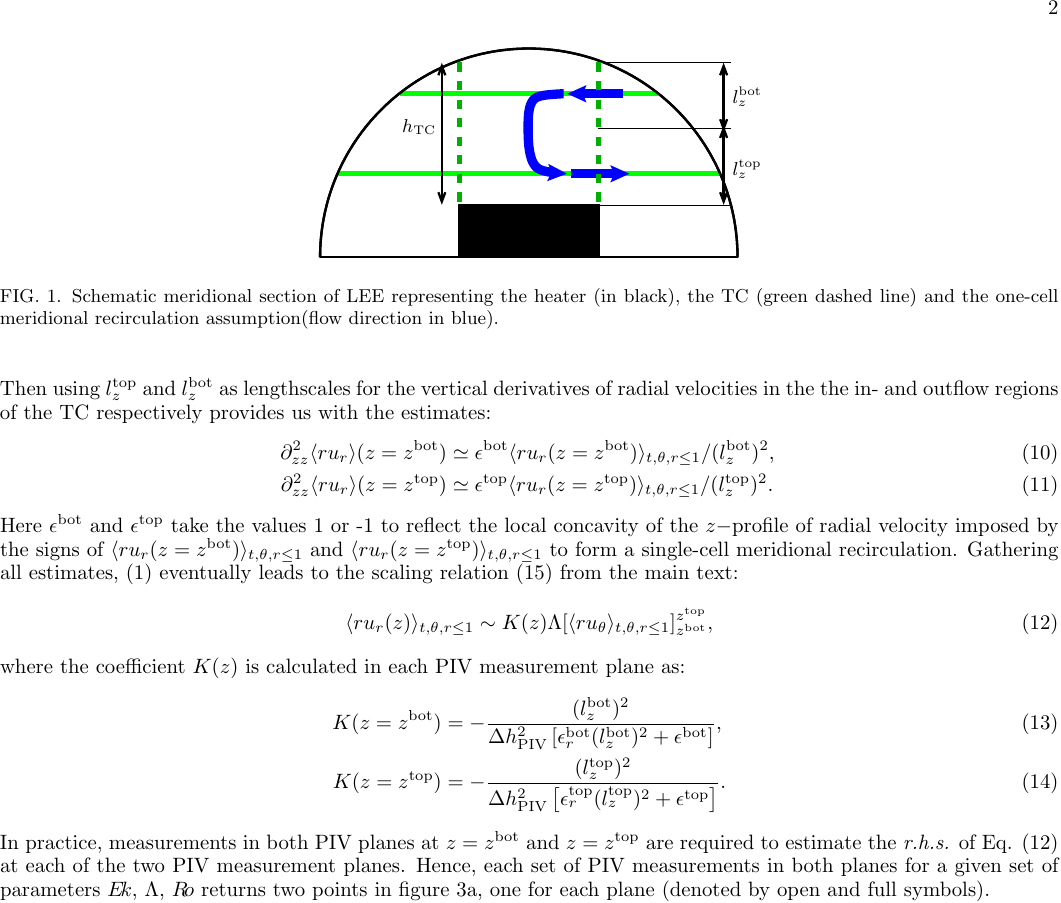}
\end{figure*}
\end{document}